\documentclass[]{piparticle-final}
\usepackage{graphicx}
\usepackage{amsmath} 

\usepackage{cite} % To compress citation ranges
\usepackage{epstopdf}

\begin{document}
\volume{7}               % To be inserted by Editor
\articlenumber{070013}   % To be inserted by Editor
\journalyear{2015}       % To be inserted by Editor
\editor{L. A. Pugnaloni}   % To be inserted by Editor
\reviewers{J. R. Darias, Universidad Sim\'on Bol\'ivar, Caracas, Venezuela.}  % To be inserted by Editor
\received{10 July 2015}     % To be inserted by Editor
\accepted{7 August 2015}   % To be inserted by Editor
\runningauthor{G. Lumay \itshape{et al.}}  % To be inserted by Editor
\doi{07.0013}         % To be inserted by Editor

\title{Flow of magnetic repelling grains in a two-dimensional silo}

% Institution references with \cite are inserted after \maketitle in theaffiliation enviroment
\author{G. Lumay,\cite{inst1}\thanks{E-mail: geoffroy.lumay@ulg.ac.be} \hspace{0.5em} 
            J. Schockmel,\cite{inst1} \hspace{0.5em}
            D. Hen\'andez-Enr\'iquez,\cite{inst2} \hspace{0.5em}
            S. Dorbolo,\cite{inst1} \hspace{0.5em}\\
            N. Vandewalle,\cite{inst1} \hspace{0.5em}
            F. Pacheco-V\'azquez\cite{inst2}\thanks{E-mail: fpacheco@ifuap.buap.mx}}

\pipabstract{
During a typical silo discharge, the material flow rate is determined by the contact forces between the grains. Here, we report an original study concerning the discharge of a two-dimensional silo filled with repelling magnetic grains. This non-contact interaction leads to a different dynamics from the one observed with conventional granular materials. We found that, although the flow rate dependence on the aperture size follows roughly the power-law with an exponent $3/2$ found in non-repulsive systems, the density and velocity profiles during the discharge are totally different. New phenomena must be taken into account. Despite the absence of contacts, clogging and intermittence were also observed for apertures smaller than a critical size determined by the effective radius of the repulsive grains.
}

\maketitle

\blfootnote{
\begin{theaffiliation}{99}
   \institution{inst1} GRASP, Physics Department B5, Universit\'e de Li\`ege, B4000-Li\`ege, Belgium.
   \institution{inst2} Instituto de F\'isica, Benem\'erita Universidad Aut\'onoma de Puebla, Apartado Postal J-48, Puebla 72570, Mexico.
\end{theaffiliation}
}

\section{Introduction}

%Intro Générale
The flow of discrete objects through an aperture in bottlenecks is an important subject in many scientific fields and for industrial applications. The objects could be animals \cite{Zuriguel2015}, pedestrians \cite{Vicsek2002,Parisi2005}, insects \cite{Parisi2012}, red blood cells \cite{Wagner2014}, bacteria \cite{Clement2013}, cars  or grains \cite{deGennes1999}. An important feature of these flows is clogging \cite{Zuriguel2014,Behringer2011,Sum2013}, which is observed during emergency escape, in traffic flows and during silo discharging. Two different flowing modes are observed: flow of contacting objects like in granular materials and flow of non contacting objects like in road traffic. Transitions between a non contacting to {a} contacting flowing mode could be observed. 

In the granular materials community, the flow of contacting grains through the output of a silo has been the subject of numerous studies \cite {Zuriguel2014,Behringer2011,Sum2013,To2001,Lozano2012,Pugnaloni2015,Dorbolo2013,Garcimartin2010,Mankoc2007,Janda2012,Geminard2012} since the seminal work of Beverloo {\it et al.} \cite{Beverloo1961}. Contrary to liquid flows, the flow of granular material at the output of a silo is constant during the discharge. Moreover, for small apertures, the probability to observe cloggings is significative. Beverloo proposed a semi-empirical relation for the granular flow rate Q (number of grains per unit time) at the output of a silo as a function of the silo aperture size D. This relation is deduced from the conjunction of two premises: (i) the flow $Q$ is blocked when the aperture is below a threshold given by $kd$, where $k$ is a free parameter and $d$ the bead diameter, (ii) the grains experience a free fall before passing through the aperture. Then,
 for the average grains speed at the silo output, one has
\begin{equation} 
\label{grainSpeed}
\langle v_{out} \rangle = \sqrt{2 g  \beta D}.
\end{equation} This relation comes from the idea that the jamming mechanism is due to the formation of a semi-circular arch before the aperture \cite{Lozano2012,Garcimartin2010}. In the case of contacting grains, this arch has a typical size proportional to the aperture, i.e., $\beta = 0.5$. 

With two-dimensional silos, the flow rate $Q$ is proportional to the mean grains speed at the outlet $\langle v_{out} \rangle$, to the mean density of grains at the outlet $\langle \rho \rangle$ and to the aperture size $D$. Therefore, the general relation for the flow rate in two-dimensional silos is
\begin{equation} 
\label{fluxVSphiVD}
Q = \langle \rho \rangle \langle v_{out} \rangle D.
\end{equation} The density of grains $\langle \rho \rangle$ is expressed in grains per surface unit. Then, the expression of Beverloo relation for two-dimentional silos is
\begin{equation}
\label{beve}
Q=\langle \rho \rangle \sqrt{2 g \beta D} (D-kd).
\end{equation}
 
%intro Felipe
It has been demonstrated that  the Beverloo's relation fails for small orifices where clogging is possible \cite{Mankoc2007}, and recently \cite{Janda2012}, it was found that the density and velocity profiles are self-similar in all the range of apertures, where the flow rate $Q$ must be modified by an exponential factor related with the lower density of material near the exit.

%Intro ce qu'on fait
Up to now, all the studies about grains flowing through apertures involved particles that interact through contact forces. In the present paper, we present experimental results concerning a repulsive granular material flowing in a two-dimensional silo. The grains are non contacting cylindrical magnets repealing each other with a highly non-linear potential. If they are not too close to each other, the magnets can be considered as magnetic dipoles. The force $\vec{F}(r,\theta)$ between two magnetic dipoles separated by a distance $r$ with an angle $\theta$ between the magnetic field direction and the line joining the dipole centers is given by:

\begin{equation}
\vec{F}(r,\theta) = \frac{3 \mu_0 m^2}{4 \pi r^4} \left [ (1- 3 \cos^2 \theta) \; \hat{r} + \sin 2 \theta \; \hat{\theta} \right],
\end{equation} where $\mu_0$ is the vacuum magnetic permittivity and $m$ is the dipolar magnetic moment. The influence of this magnetic interaction on the flow of dense granular materials has been already considered in a previous study \cite{Lumay2008}. In the present configuration (see set-up section), $\theta = 90^\circ$. Therefore, the non-linear repulsive force strength between two grains in the silo is $F(r) = \frac{3 \mu_0 m^2}{4 \pi r^4}$.  Considering this repulsive interaction, we investigate the flow rate at the silo output for different aperture sizes. Moreover, the relations between the flow rate, the velocity profile along the silo aperture and the density behind the aperture are investigated.

%%%%%%%%%%%%%%%%%%%%%%%%%%%%%%%%%%%%%%%%%%%%%%%%%%%%%%%%%%%%%%%%%%%%%%%%%%%%%%%%%%%%%%%%%%%
%Experimental setup
%%%%%%%%%%%%%%%%%%%%%%%%%%%%%%%%%%%%%%%%%%%%%%%%%%%%%%%%%%%%%%%%%%%%%%%%%%%%%%%%%%%%%%%%%%%

\section{Set-up}

\begin{figure}
  \includegraphics[width=8.5cm]{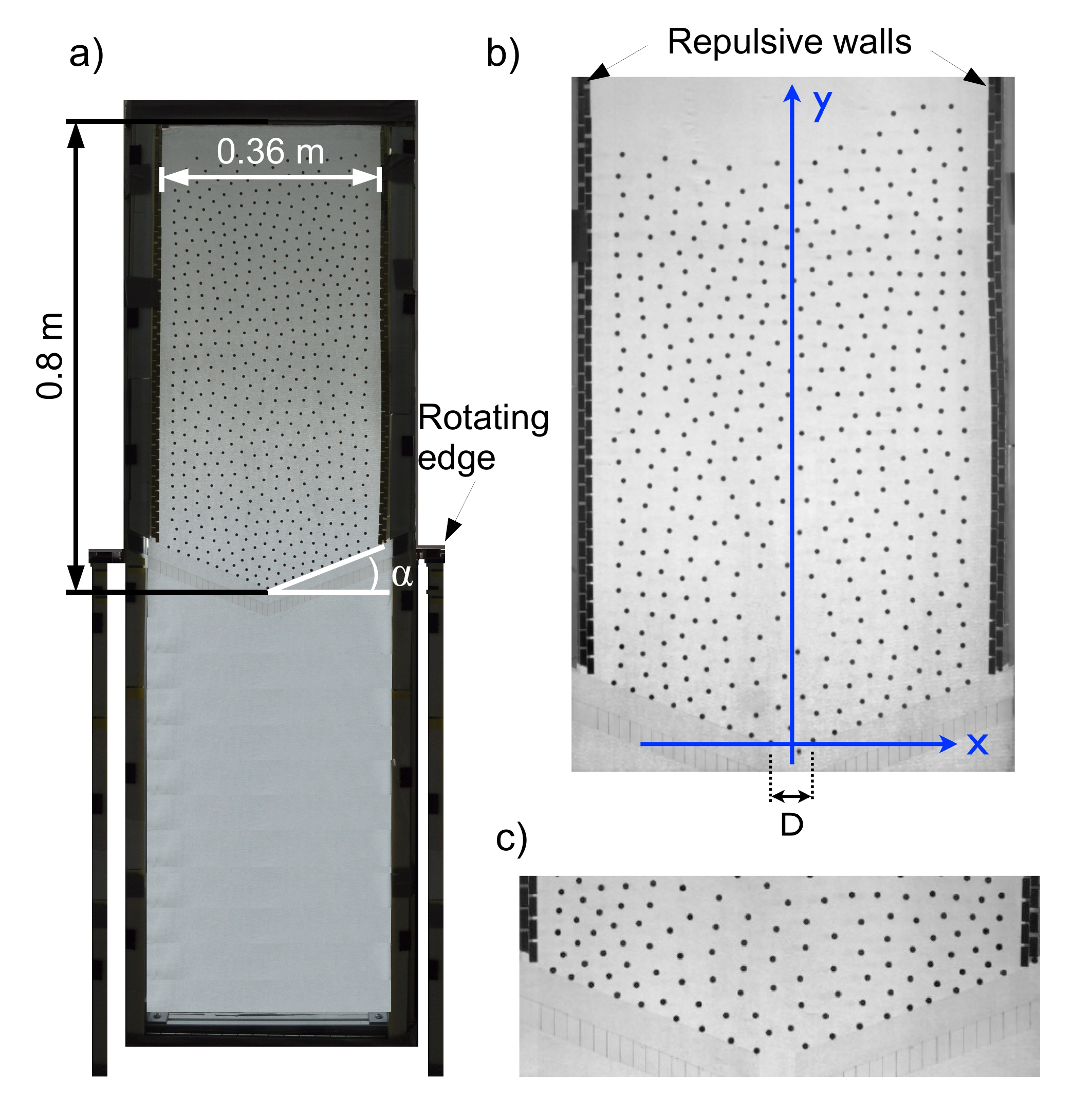}
  \caption{(a) and (b) Annotated picture of the two-dimensional silo filled with magnets in repulsive configuration. (c) Magnetic arch observed sometimes for small silo aperture sizes.}
\label{fig:SetUp}
\end{figure}

A sketch of the experimental set-up is presented in Fig. \ref{fig:SetUp} (a). A set of $N_{tot}=565$ Neodymium disc magnets (thickness  $t=3.00 \pm 0.02$ mm, diameter $d=5.00 \pm 0.02$ mm, mass $m=0.4$ g and surface magnetic field $H=5.9$ kG) was introduced in a two-dimensional vertical silo. The frontal walls of the silo consisted of two transparent glass plates separated by $3.10 \pm 0.05$ mm and the lateral walls were composed of Neodymium bars with their magnetization oriented perpendicularly to the glass walls. The grains were carefully introduced with the same orientation in order to obtain repulsive grain-grain and grain-wall interactions [Fig. \ref{fig:SetUp} (b)]. The silo was divided in two compartments separated by a movable gate. The structure can be rotated about an horizontal axis in order to reset the silo. The gates inclination $\alpha = 22^\circ$ ensured the complete discharge of the 565 magnets.

Once the upper compartment was charged, the aperture size $D$ was fixed with the flow initially blocked by a electromechanic switch. When the switch was turned off, the grains started to flow and the discharge was filmed with a high speed camera at 250 fps. Because the particles were non-contacting, they were easily tracked using ImageJ in order to obtain their positions, velocities and the flow rate. 

%%%%%%%%%%%%%%%%%%%%%%%%%%%%%%%%%%%%%%%%%%%%%%%%%%%%%%%%%%%%%%%%%%%%%%%%%%%%%%%%%%%%%%%%%%%
%Results
%%%%%%%%%%%%%%%%%%%%%%%%%%%%%%%%%%%%%%%%%%%%%%%%%%%%%%%%%%%%%%%%%%%%%%%%%%%%%%%%%%%%%%%%%%%

\section{Results}

Three videos \cite{videos} show the magnetic grains flow in the cases $D=0.03$ m, $D=0.08$ m and $D=0.18$ m. For  $D=0.03$ m; we observe an intermittent flow and the formation of an arch. These magnetic aches [see Fig. \ref{fig:SetUp} (c)] are formed for small aperture size ($D < 0.04$ m) and are able to block the flow. Before a magnetic arch formation, a continuous flow is observed. Moreover, as with classical granular materials \cite{Lozano2012}, an arch can be broken with vibrations to recover a continuous flow. The analysis of arches formation probability is outside the scope of the present study. Therefore, the analysis of the results is made for a larger aperture size ($D \geq 0.04$ m) and considering that $k=0$. 

\begin{figure}
	\includegraphics[scale=0.6]{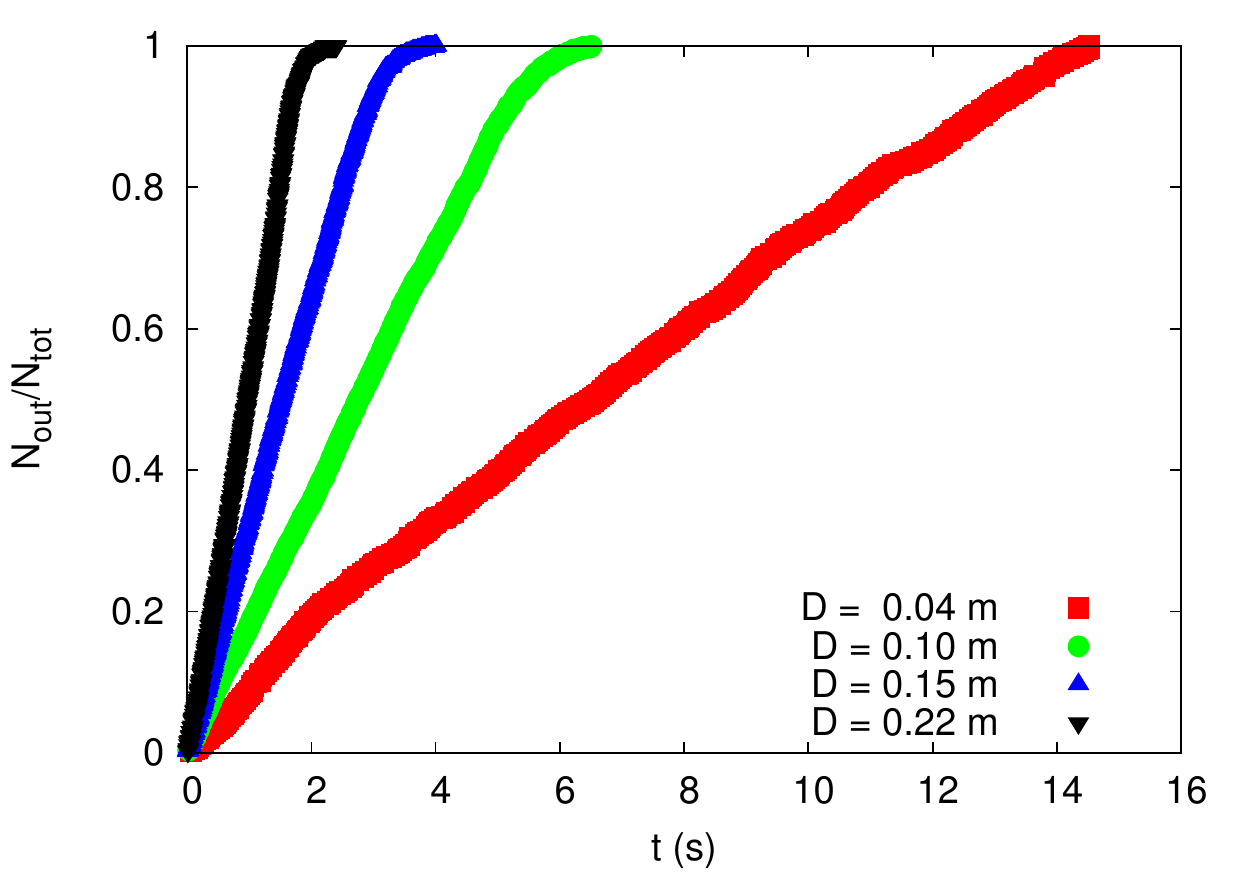}
	\caption{Fraction of grains escaped from the silo $N_{out}/N_{tot}$ as a function of time for different aperture sizes $D$.}
	\label{fig:N_out}
\end{figure}

\begin{figure}
	\includegraphics[scale=0.6]{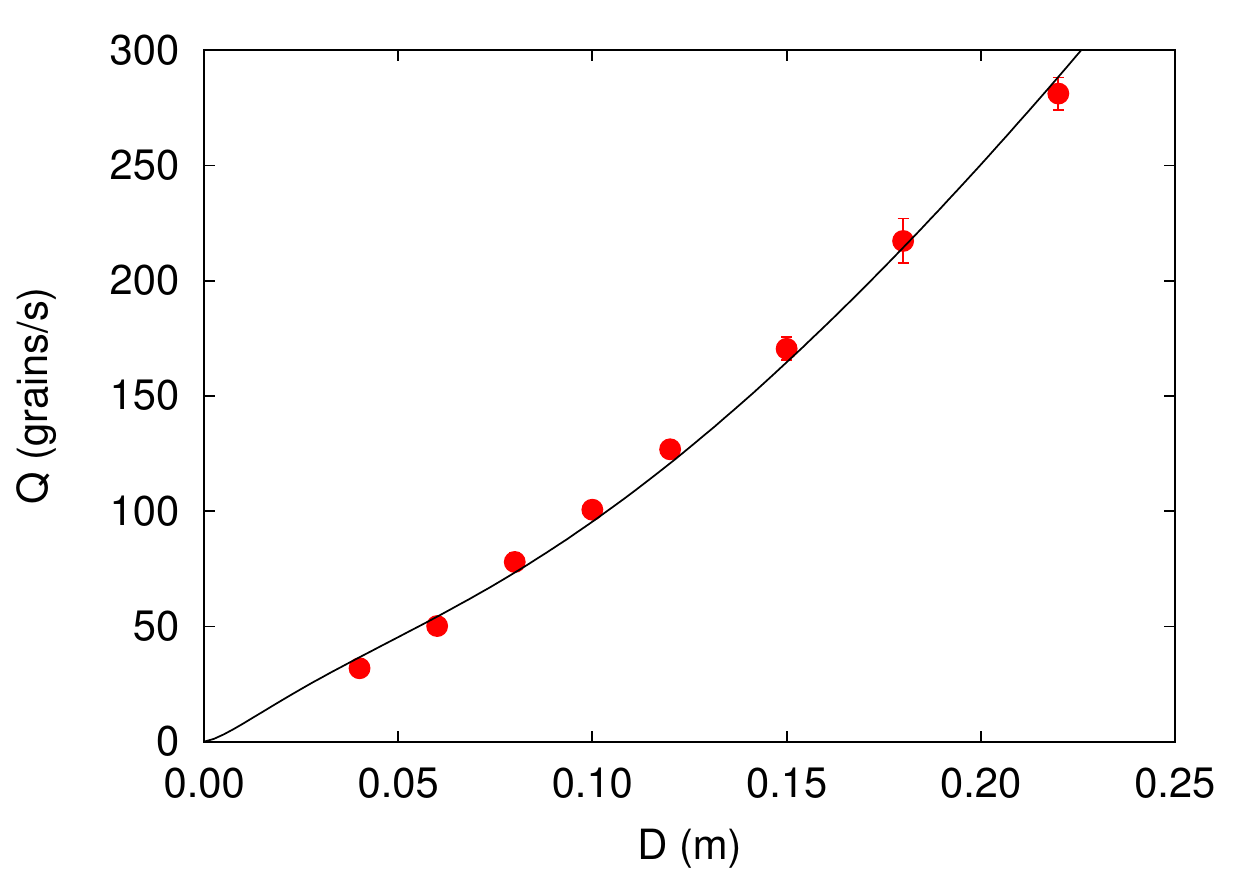}
	\caption{Flow rate $Q$ as a function of the silo aperture size $D$. The continuous line corresponds to the Beverloo relation with an additional exponential factor [see Eq. (\ref{modifBeverloo})]. Each point corresponds to an average over 3 experiments. Error bars are indicated.}
	\label{fig:flowRate}
\end{figure}

%flow rate analysis
Figure \ref{fig:N_out} shows the fraction of grains escaped from the silo $N_{out}/N_{tot}$ as a function of time for different aperture sizes $D$. The curves are roughly linear at the exception of the end of the emptying process. Therefore, as with contacting grains, the flow rate is constant during the silo discharge. The beginning of each curve has been fitted by a linear law to obtain the flow rate $Q$. The evolution of the flow rate $Q$ (expressed in grains per second) as a function of the aperture size D is presented in Fig. \ref{fig:flowRate}. This flow rate can be fitted with a classical Beverloo relation [Eq. (\ref{beve})] for big orifices. However, the fit fails for small ones. To fit the flow rate over the whole range of orifice sizes, an exponential factor has been added. In order to investigate the origin of this exponential factor, the grains speed and the grains density at the silo output have been analyzed. 

\begin{figure}
	\includegraphics[scale=0.6]{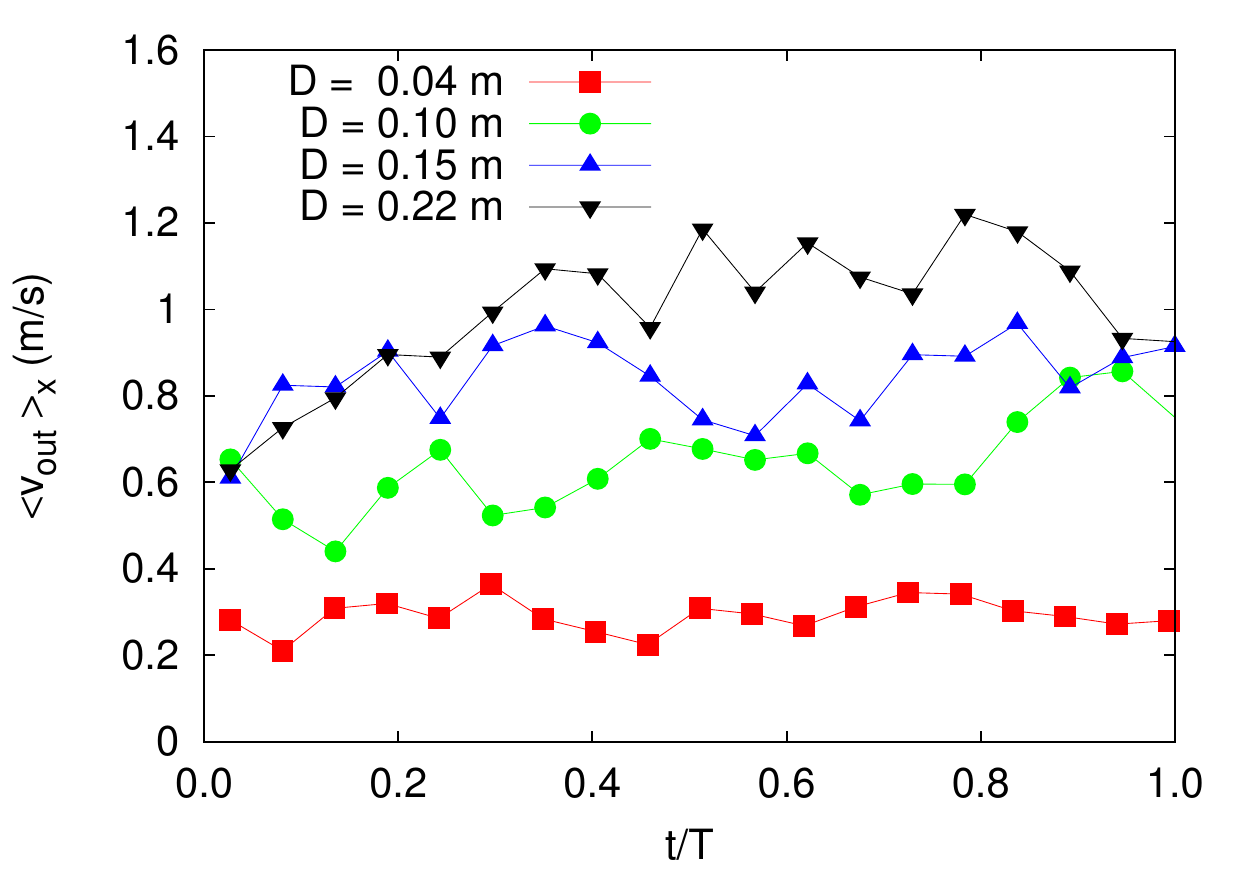}
	\caption{Temporal evolution of grains velocity averaged over the silo aperture $\langle v \rangle_x$ for different aperture sizes $D$. The time is normalized by the silo discharging time $T$.}
	\label{fig:timeSpeed}
\end{figure}

%Grains speed analysis
From the tracking of grain trajectories during the discharge, grains velocity at the silo output $v_{out}$ has been measured. The temporal evolution of the grains velocity $\langle v_{out} \rangle_x$ at the silo output is presented in Fig. \ref{fig:timeSpeed}. The notation $\langle \rangle_x$ corresponds to an averaging over the aperture. The time is normalized by the silo total discharging time $T$. For small aperture sizes ($D < 0.15$ m), the output velocity is roughly constant during the discharge at the exception of the end of the process. However, for a larger aperture size ($D > 0.15$ m), the output velocity increases with the time. This increase is certainly due to the finite size of the silo. Indeed, when the aperture size is approaching the silo width, the confinement at the output is less important and the whole bulk is experiencing a free fall.

\begin{figure}
	\includegraphics[scale=0.6]{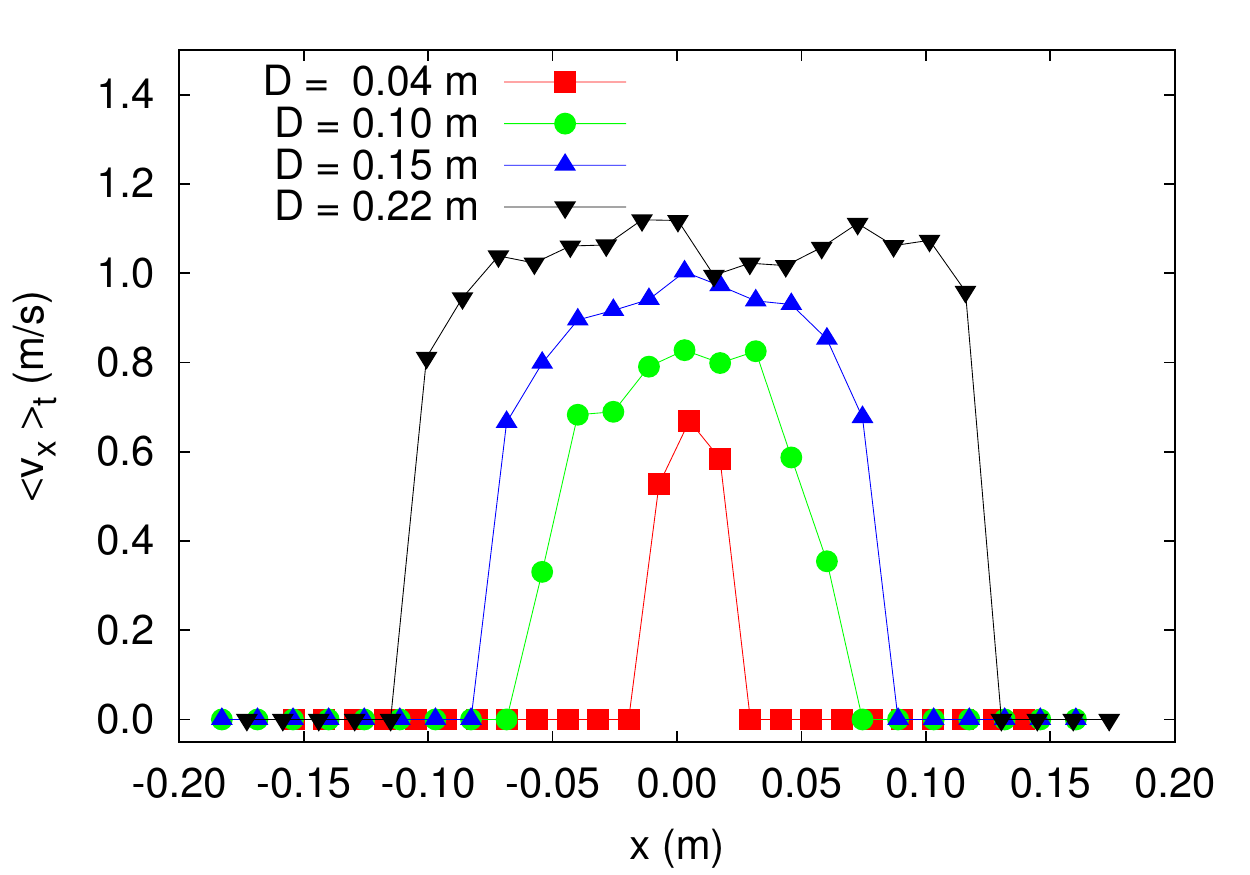}
	\caption{Grain velocity profile $\langle v \rangle_t$ along the silo aperture for different silo aperture sizes $D$.}
	\label{fig:speedX}
\end{figure}

Figure \ref{fig:speedX} shows the grain velocity profile along the silo aperture. The velocity is averaged over the whole discharging time $T$. An important velocity gradient is observed near the borders of the aperture. Apart from this zone, a plug flow is observed. This plug flow constitutes a strong differentiation with classical granular flows \cite{Janda2012}. 

\begin{figure}
	\includegraphics[scale=0.6]{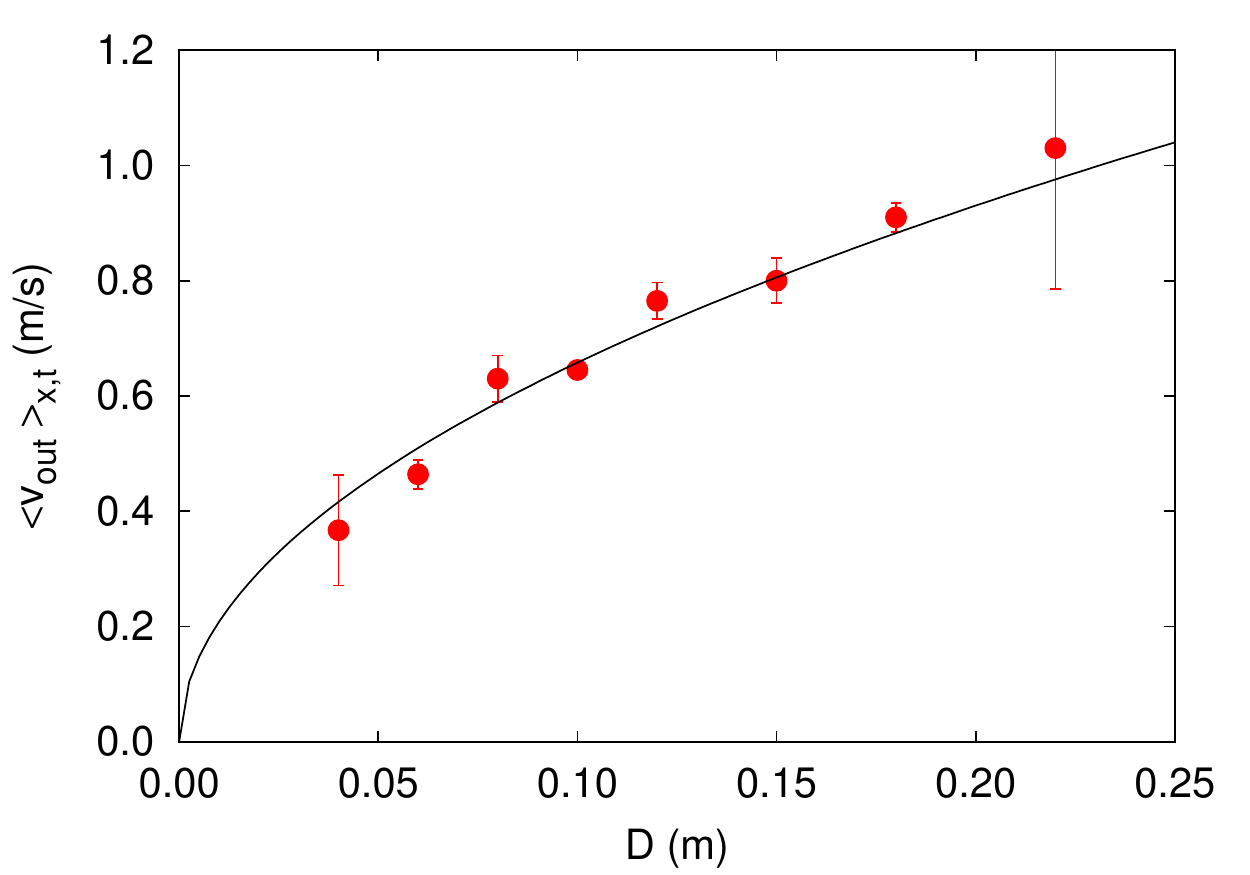}
	\caption{Evolution of grain velocity $\langle v \rangle_{x,t}$ at the output of the silo as a function of $D$. The continuous line corresponds to a fit with Eq. (\ref{grainSpeed}). Each point corresponds to an average over 3 experiments. Error bars are indicated.}
	\label{fig:meanSpeedVSdiameter}
\end{figure}

The evolution of the grain velocity $\langle v \rangle_{x,t}$ at the output of the silo as a function of $D$ is presented in Fig. \ref{fig:meanSpeedVSdiameter}. The notation $\langle \rangle_{x,t}$ corresponds to an averaging over the aperture and over the whole discharging time. The increase of the velocity with $D$ is well fitted by the square root law Eq. (\ref{grainSpeed}), with $\beta = 0.220 \pm 0.002$. In granular materials, this equation gives the grains speed corresponding to a free fall from a height $\beta D$, where $\beta$ is a geometrical factor equal to $0.5$. With repealing magnetic grains, the association of the plug flow and the free fall with $\beta = 0.22$ allows to conclude that the grains  experience a free fall in a rectangular region situated above the aperture. The height of this rectangular region is $0.22 D$.

\begin{figure}[t]
	\includegraphics[scale=0.6]{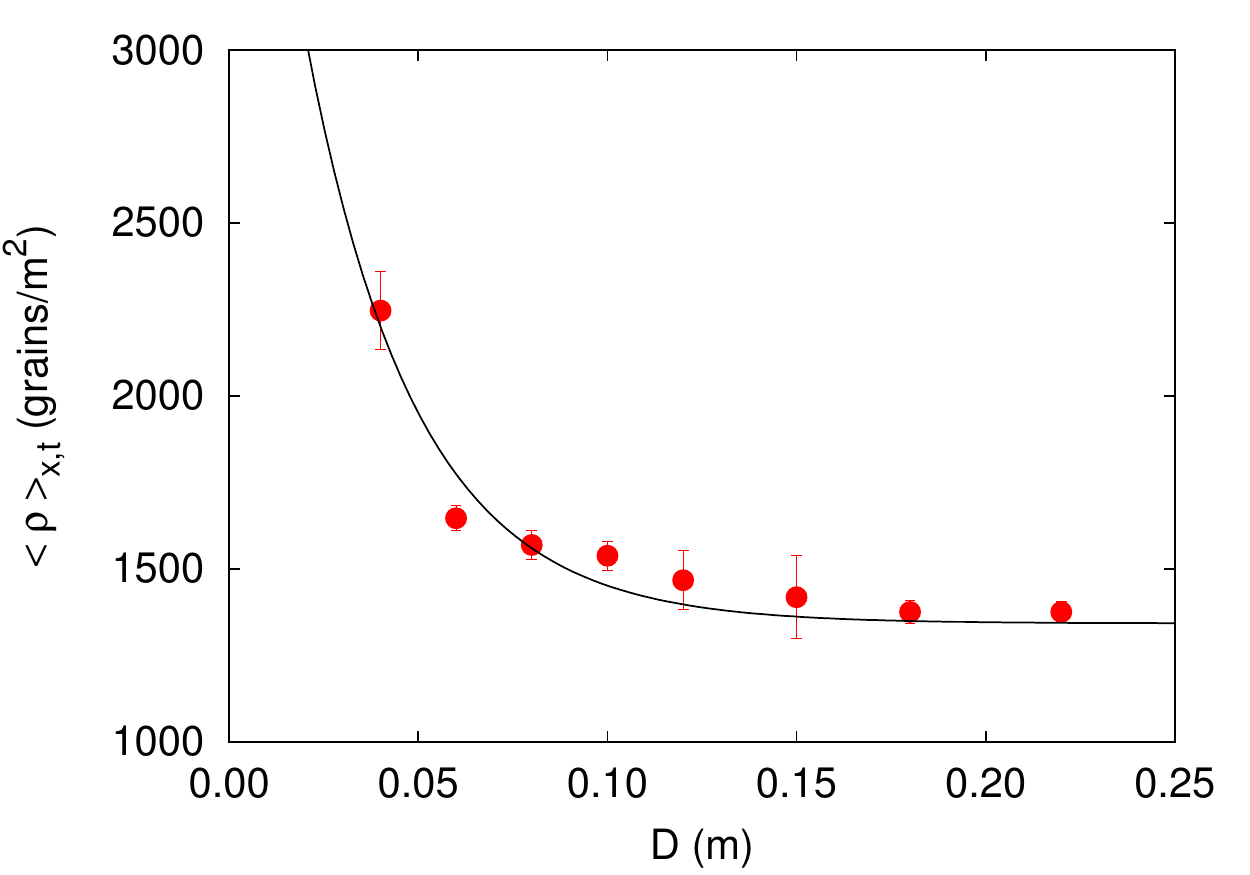}
	\caption{Averaged density $\langle \rho \rangle_{x,t}$ above the silo aperture  as a function of the silo aperture. The continuous line corresponds to a fit with Eq. (\ref{phi}). Each point corresponds to an average over 3 experiments. Error bars are indicated.}
	\label{fig:densityVSD}
\end{figure}

%grains density analysis
The density of grains $\langle \rho \rangle_x$ just above the output is the number of beads whose center of mass is included in the measurement box divided by the box surface. The measurement box is a rectangle of width $D$ and height $10 d$ situated just behind the aperture. The density temporal evolution is plotted in Fig. \ref{fig:densityVSt} for different values of $D$. At early time, a quick decrease of the density is observed, then the density reaches a plateau. Finally, near 70 \% of the total time of the discharge, the density decreases until the silo is empty. The time averaged density $\langle \rho \rangle_{x,t}$ is evaluated during the constant regime. Figure \ref{fig:densityVSD} shows the time averaged density $\langle \rho \rangle_{x,t}$ above the output as a function of the aperture size $D$. The density is found to decrease with the aperture size. This decrease constitutes also a differentiation with classical granular flows \cite{Janda2012}. The decrease of the 
density as a function of the aperture size is well fitted by the decreasing exponential law:

\begin{figure}[t]
	\includegraphics[scale=0.6]{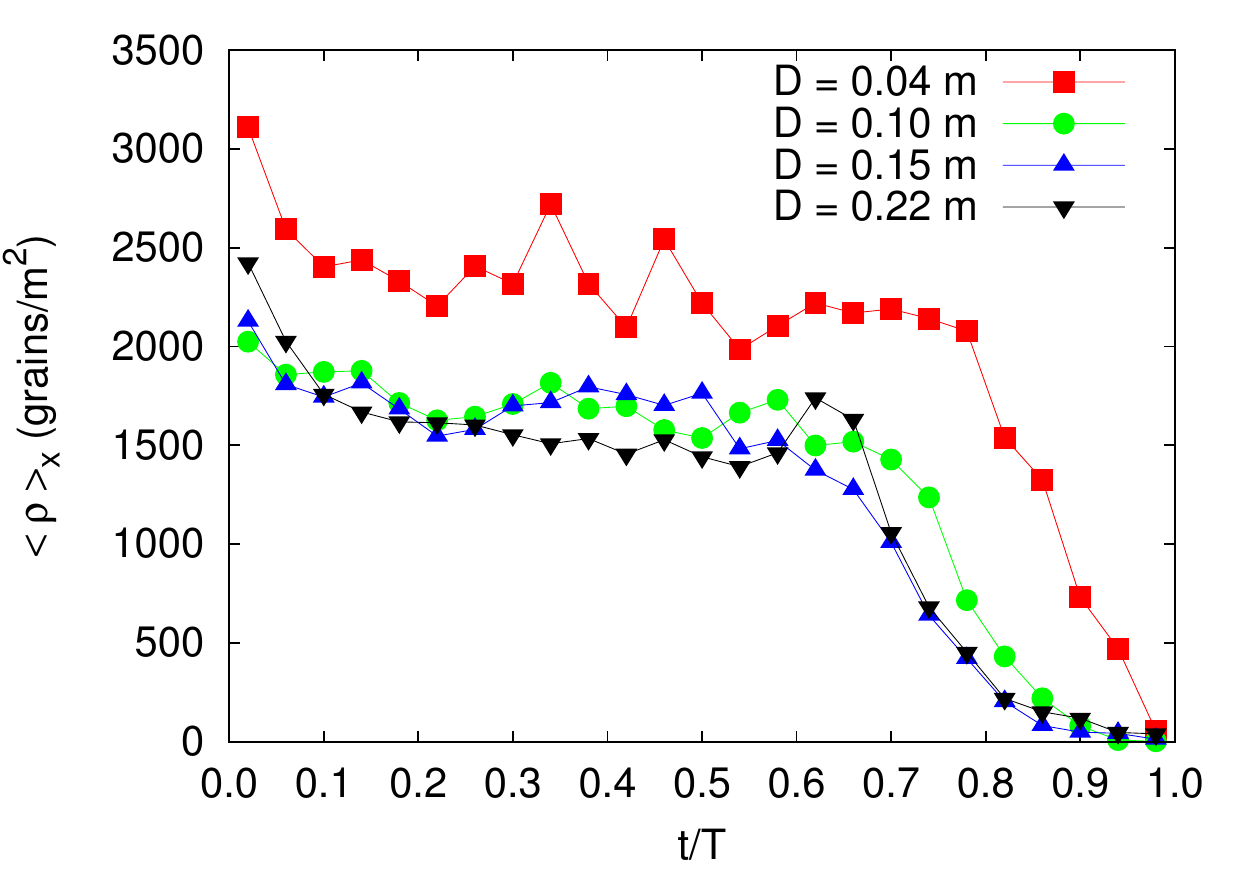}
	\caption{Temporal evolution of the density $\langle \rho \rangle_{x}$ averaged over the aperture. The measurement box is a rectangle of width $D$ and height $10 d$ situated just behind the aperture. }
	\label{fig:densityVSt}
\end{figure}

\begin{equation}
\label{phi}
\langle \rho \rangle_{x,t} = \rho_{\infty} + (\rho_0-\rho_{\infty}) e^{-\frac{D}{\alpha}},
\end{equation} with $\rho_{\infty}  = 1342 \pm 29$ grains/m$^2$, $\rho_0 = 4737 \pm 716$ grains/m$^2$ and $\alpha = 0.029 \pm 0.004$ m. By combining Eqs. (\ref{grainSpeed}), (\ref{fluxVSphiVD}) and (\ref{phi}), we obtain a modified version of Beverloo relation:

\begin{equation} 
\label{modifBeverloo}
Q = \left[ \rho_{\infty} + (\rho_0-\rho_{\infty}) e^{-\frac{D}{\alpha}} \right] \sqrt{2 g  \beta D} D.
\end{equation} This relation is plotted in Fig. \ref{fig:flowRate} with the parameters $\rho_0$, $\rho_{\infty}$, $\alpha$ and $\beta$ obtained previously from the fit of the average grain speed and of the density as a function of the aperture size $D$. 

\section{Conclusion}

The flow of repealing magnetic grains at the output of a two dimensional silo has been analyzed experimentally. A plug flow has been observed at the silo output with a grain speed corresponding to a free fall in a rectangular region situated above the output. Comparing to the flow of non-magnetic contacting grains in the same geometry, the observed plug flow constitutes a difference. However, as in the contacting grains case, the flow is constant during the silo discharge. The density of grains measured just above the silo output is found to decrease exponentially with the silo aperture size. This decrease of the density constitues a second difference comparing to the contacting grains case. Finally, the analysis of both grains speed and bulk density at the silo output allows to propose a modified Beverloo relation that fit the flow rate as a function of the aperture size. This modified Beverloo relation is the classical power-law with an exponent $3/2$ associated to a decreasing exponential. 

The approach used during the last decades to better understand the complex behavior of granular material is an inspiration source for scientists studying crowd dynamics, traffic jams, ethology. The set of magnetic repealing grains considered in the present study could be also an interesting model system to be compared with crowds, cars and animals. Indeed, contrary to classical granular materials, the set of repealing magnets are non-contacting. Moreover, the magnetic interaction between the magnets is well known. 

\begin{acknowledgements}
F.P.V and D.H.E. thank Conacyt Mexico for financial support through the program ``Fondo Sectorial de Investigaci\'on para la Educaci\'on''. G.L and J.S. thank FNRS (Grant PDR T.0043.14) and University of Li\`ege (Starting Grant C-13/88).  
\end{acknowledgements}

\end{document}